\documentclass[journal]{IEEEtran}

\usepackage{graphicx}
\usepackage{amsmath,amssymb}
\usepackage{siunitx}
\usepackage{array}
\usepackage{url}
\usepackage{booktabs}
\usepackage{hyperref}
\usepackage{xcolor}
\usepackage{cite}
\hypersetup{hidelinks}
\usepackage{float}
\usepackage[english]{babel}
\usepackage[autostyle, english = american]{csquotes}
\MakeOuterQuote{"}
\usepackage{amsmath} % Make sure this is in your preamble

\title{Quench Protection in Insulated REBCO Conductors: Design Optimization and Fast Detection via REBCO SQD}

\author{\underline{Hajar~Zgour}, Walid~Abdel~Maksoud, Bertrand~Baudouy, Antoine Guinet \\% 
	\thanks{Authors are with CEA Paris-Saclay, Universit\'e Paris-Saclay (corresponding author: \texttt{hajar.zgour@cea.fr}).\newline Funding: France 2030 / ANR-24-EXSF-0006.}}
\begin{document}
	\maketitle

	\begin{abstract}
		This work was conducted within the framework of the exploratory French project PEPR SupraFusion, which aims to advance the field of fusion energy by developing High-Temperature Superconductor (HTS)-based demonstrators capable of storing significant energy while operating under high magnetic fields and currents.
		Ensuring a reliable protection during a quench in Insulated REBCO conductors is challenging : slow normal-zone propagation and validation delays allow the hotspot's temperature to reach damaging levels. We compare (i) conductor protection via copper-stabilizer optimization and (ii) a co-wound,  REBCO superconducting quench detector (SQD) that is electrically isolated yet thermally coupled and intentionally deoxygenated to lower $T_c$ and $I_c$ for an earlier transition. One-dimensional THEA modeling shows that a good choice of stabilizer cross-section makes the protection possible during quench events by keeping the temperature of the hotspot within a safe limit. The simulations also demonstrate that the use of a REBCO SQD enables the quench detection at lower temperatures. 
	\end{abstract}


	\section{Introduction}
	\IEEEPARstart{W}{ithin} the PEPR \textit{SupraFusion} program (France~2030), we address quench detection and protection in insulated REBCO stacks for fusion-relevant magnets. In such windings, slow normal-zone propagation velocities of a few $\mathbf{cm.s^{-1}}$,  and the validation time needed to avoid false triggers delay voltage-tap detection and can raise hotspot temperatures to damaging levels. Our goal is to make insulated stacks \emph{protectable} by accelerating the voltage detection time without compromising the protection or adding complexity to the design or integration.
	
	We investigate two complementary strategies. First, \emph{stabilizer optimization}: selecting the copper cross-section to balance detectability against Joule heating. Second, \emph{Superconductor Quench Detector (SQD)-assisted detection}: co-winding a lightly stabilized REBCO tape, which produces a higher voltage than the conductor for the same disturbances. These two methods will be tested on stack-like conductors in the H0 facility at CEA-Saclay. To prepare this experimental proof of concept, we use the THEA code by Cryosoft to model the conductors' behavior during quench events \cite{THEA24}. The 1-D electro-thermal models in THEA (conductor-only and conductor+SQD variants) are constrained by the test facility's available current, temperature and magnetic field conditions. They allowed to study the influence of the stabilizer cross section, the SQD's operating current and degradation factor on the conductor's hotspot during a quench.

	\section{Quench Protection Concepts}

	\begin{figure}[t]
		\centering
		\begin{minipage}[t]{0.48\textwidth}
			\centering
			\includegraphics[width=\linewidth]{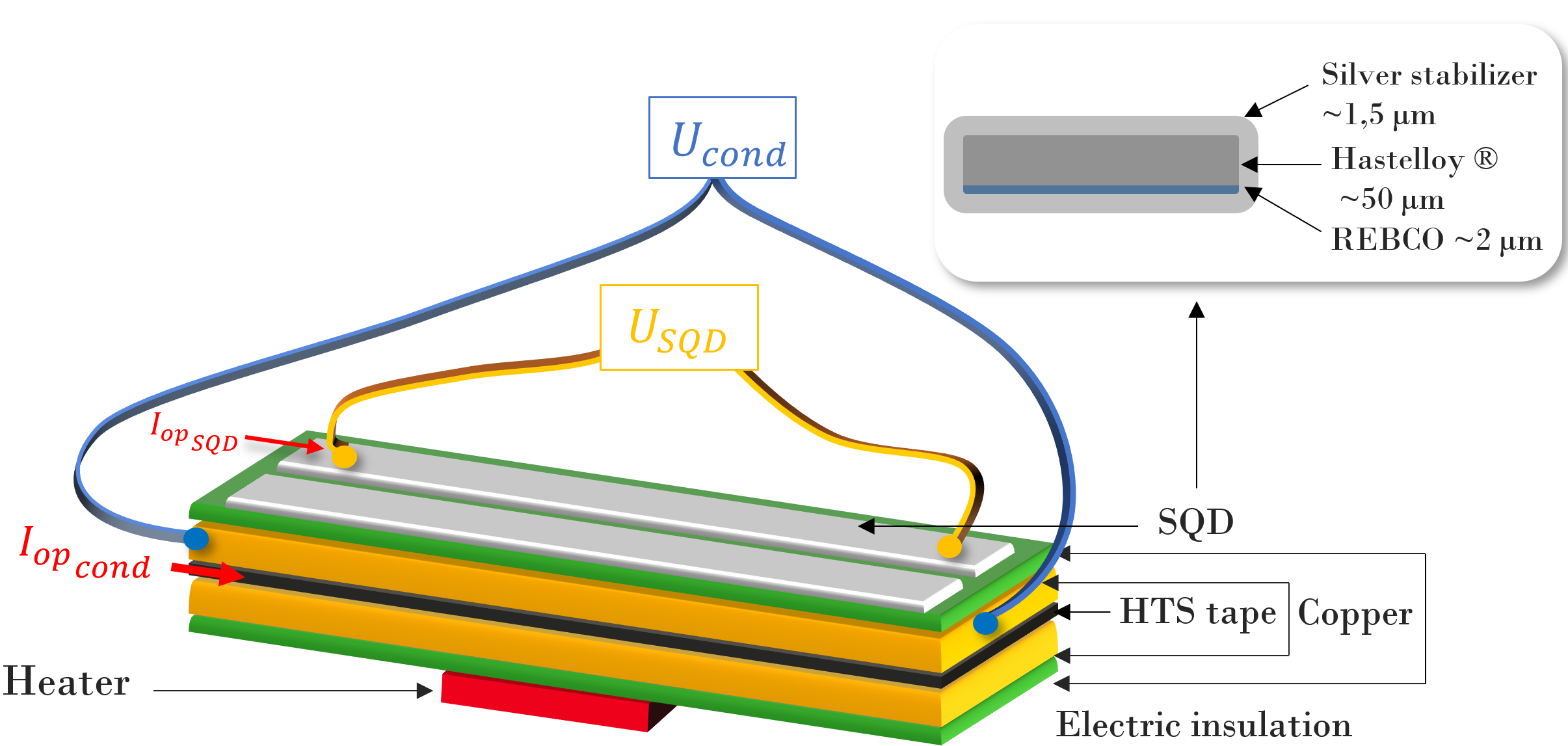}
			\caption{Conceptual layout of the co-wound sensor. A lightly stabilized REBCO SQD (top) is thermally coupled to the insulated REBCO conductor (bottom) and electrically isolated from it. Independent voltage taps read $U_{\mathrm{cond}}$ and $U_{\mathrm{SQD}}$. The conductor carries the operating current; the SQD is biased at a small DC current $I_{\mathrm{op,SQD}}$ to enhance its resistive response. A local heater provides a controlled disturbance.}
			\label{fig:concept_layout}
		\end{minipage}\hfill
		\begin{minipage}[!h]{0.48\textwidth}
			\centering
			\includegraphics[width=\linewidth]{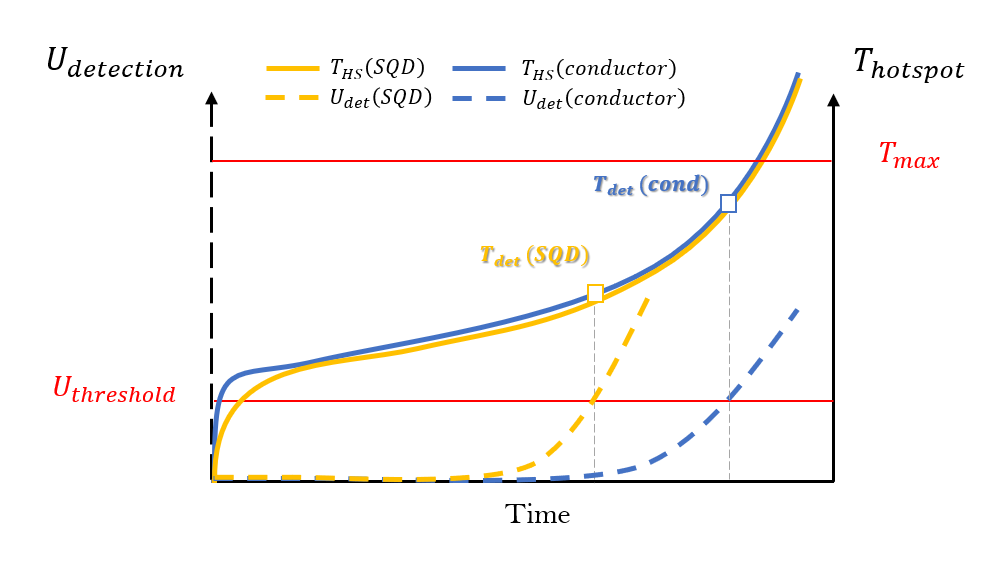}
			\caption{Sensing principle. Solid lines: hotspot temperatures of the conductor and the SQD; dashed lines: corresponding voltages over a fixed gauge length. The SQD reaches the voltage threshold $U_{\text{threshold}}$ earlier ($t_{\mathrm{det}}(\mathrm{SQD})$) than the conductor ($t_{\mathrm{det}}(\mathrm{cond})$), yielding a cooler hotspot at detection and after the validation delay, relative to the allowable limit $T_{\max}$.}
			\label{fig:concept_signals}
		\end{minipage}
	\end{figure}
	
	\subsection{Protection With Stabilizer}
	During an adiabatic quench (detection phase + exponential current dump into an external resistor), the conductor's hotspot temperature ($T_{HS}$) evolution can be expressed by Eq.(\ref{eq:integral_HS}), where $T_{HS}$ and $T_{op}$ are the hotspot and the operating temperatures respectively, and $T_{det}$ is the hotspot's temperature when the voltage threshold is crossed.

	$$T_{HS} - T_{op} =  (T_{det} - T_{op}) + (T_{HS} - T_{det})  $$
	
	\begin{equation}
		T_{HS}= \underbrace{\int_{0}^{t_{det}}\frac{\rho_{\mathrm{e}} I_{op}^2}{\rho_{m} c_{p_{\mathrm{Cu}}} S^2_{Cu}} dt}_{\displaystyle \mathrm{Detection \; phase}} \;+\; \underbrace{ \int_{t_{det}}^{\inf}\frac{\rho_{\mathrm{e}} I_{op}^2}{\rho_{m} c_{p_{\mathrm{Cu}}} S^2_{Cu}} e^{\frac{-2t}{\tau}} dt}_{\displaystyle \mathrm{dump \; phase}} + T_{op}
		\label{eq:integral_HS}
	\end{equation}
	
	\noindent  where $S_{\mathrm{Cu}}$ is the copper stabilizer cross-section (considered dominant compared to the other layers: Hastelloy and REBCO); $\rho_{\mathrm{e}}$ and $\rho_{\mathrm{m}}$ are the copper
	resistivity and density respectively; $c_{\mathrm{p_{Cu}}}$ is the copper's heat capacity per unit length; $t_{\mathrm{det}}$ is the
	detection time; and $\tau$ is the electrical time constant of
	the current dump. For more details on the hotspot temperature formulations, please refer to \cite{Iwasa2009_Protection}.

	Here, the direct influence of the copper cross section can be easily observed: thicker copper increases the heat capacity term and hence reduces the hotspot temperatures. However, the key parameter $t_{det}$ (quench detection time) also strongly depends on the copper content via the quench propagation velocity $v_q$. In Equation \ref{eq:t_det}, $L_{det}$ is the normal zone length when the quench is detected.
	
	\begin{equation}
		t_{det} = \frac{L_{det}}{v_q} = \frac{U_{\mathrm{threshold}} \ S_{Cu}}{I_{op} \ \rho_{\mathrm{e}} \  v_q}
		\label{eq:t_det}
	\end{equation}
	where the propagation velocity is expressed, in the case of a stack-like copper stabilized conductor \cite{Iwasa2009_Protection} :

	\begin{align}
		v_q &= \nonumber \\
		&\quad \sqrt{ \frac{
				J_{op}^2 \, \rho_e \, k
			}{
				\left( 
				S_{Cu} c_{p_{\mathrm{Cu}}}
				- \frac{1}{k} \frac{dk}{dT}
				\int_{T_{op}}^{T_c} S_{Cu} c_{p_{\mathrm{Cu}}} \, dT
				\right)
				\int_{T_{op}}^{T_c} S_{Cu} c_{p_{\mathrm{Cu}}} \, dT
		} }
	\end{align}

	Where $k$ is the equivalent thermal conductivity and $T_c$ the critical temperature of the conductor. Consequently, adding more stabilizer to the conductor slows down the quench propagation velocities, causing higher hotspot temperatures. However, a direct correlation between the stabilizer content and the hotspot temperature can not be intuitively guessed because the variables are inter-dependent. Hence, a parametric study is done numerically and is to be carried out experimentally to find the appropriate copper thicknesses that balance the quench detectability and the Joule heating in our conductors.

	%%%%%%%%%%%%%%%%%%%%
	%%%%%%%%%%%%%%%%%%%%
	%%%%%%%%%%%%%%%%%%%%

	\subsection{Detection With a REBCO SQD}

	Because the intrinsic voltage generated by an insulated REBCO conductor during a quench is small, we suggest to co-wind a dedicated SQD : a relatively narrow REBCO tape which respects the following criteria (Fig.\ref{fig:concept_layout}):
	\begin{itemize}[]
		\item Slight electrical stabilization $S_{Cu} \approx 0$ : with a minimal stabilizer cross section (Silver in our case) up to 1000 times lower than that of the conductor, the SQD's electrical resistance is expected to be higher, allowing the detection of higher voltages.
		
		\item $T_{cs, SQD} \ge T_{cs,cond} $: the quench of the conductor causes the transition to the normal state of the SQD and not inversely.
		
		\item Good electrical insulation: the SQD is operating thanks to an independent DC current source (delivering up to 20 A), and its voltage is measured independently. Hence, an electrical insulation is ensured thanks to the high dielectric strength of glass epoxy, along with the relatively low voltages of the power sources $U_{max} = 10 \, \mathrm{V}$. 
		
		\item Bifilar layout : the SQD tapes are implemented in a bifilar setup to reduce loop area and cancel potential inductive voltages during current ramps and dump. Thereby improving measurement accuracy and limiting false triggers.
		
		\item Good thermal coupling: when the conductor transits and its temperature rises, the heat must flow rapidly to the SQD to trigger its quench. The shorter is the delay, the faster the detection. Moreover, the low thermal resistance allows the heat due to Joule heating in the SQD to be dissipated into the conductor, avoiding its damage.

		\item $T_{c,SQD} > 20 \ \mathbf{K}$ and $B_{c,SQD} > 20 \  \mathbf{T}$: for it to be used in fusion magnets applications, the detector must be superconducting at relatively high temperatures and magnetic fields, hence the choice of a REBCO tape instead of an LTS wire.
		
		\item Higher detection sensitivity: first, in order to produce high voltages as soon as the conductor's temperature exceeds its $T_{cs, cond}$ by several Kelvins, the SQD must have a relatively low critical temperature $T_c$ to narrow down the current sharing phase. Second, the critical current $I_{c,SQD}$ must be lowered to enable operating currents $I_{op,SQD}$ of tens of Ampers without compromising the voltage sensitivity. This "tuning" of the critical properties of the SQD dedicated tapes can be achieved by heat treatment. Indeed, when REBCO tapes are exposed to high temperatures ($> 230 \ \mathrm{K}$) for relatively long durations (several hours) and under a controlled atmosphere, the oxygen in the superconducting layer diffuses out, causing degradation of the superconducting properties. Deoxygenation of REBCO tapes has been studied in literature in the aim of avoiding tape damage during manufacturing processes and has been proven to be controllable and reproducible \cite{Pan2023,Lu2021,Cayado2023,Stangl2019PhD,Peng2024,Bradford2024}. However, in our case, this degradation is \textit{intentional} and would allow us to tailor the tapes for our dedicated SQD. It is worth noting that without this degradation, the SQD would behave as a standard tape, with the same voltage responses and the same sensitivity. \\

	\end{itemize} 
		The experimental proof of concept of these quench detectors is under preparation. Several heat treatments have been carried out to find the targeted degradations. Results of the deoxygenation study and the proof of concept will be published elsewhere.

		\smallskip

	With these conditions, we aim to reach the threshold ($U_{\text{threshold}}$) \emph{earlier} than the conductor voltage (\,$t_{\mathrm{det,SQD}}<t_{\mathrm{det,cond}}$\,). This earlier trigger translates into a \emph{lower} conductor hotspot at detection $T_{\mathrm{det,SQD}} < T_{\mathrm{det,cond}}  $,  and after validation, a greater margin with respect to $T_{\max}$ (see Fig.~\ref{fig:concept_signals}).\\

	\section{Numerical proof of concept}
	
	\subsection{Operating conditions}
	The proof-of-concept will be carried out in the H0 test facility at CEA\cite{Avronsart2019}. The background field is provided by the H0 magnet, a large-bore NbTi system operated in liquid helium that can deliver up to 3 T with high homogeneity  and a clear bore of about 480 mm. The sample holder itself is conduction-cooled and powered via a 2 m long current lead.
	
	\smallskip
	Within this facility, we will test stack-like insulated REBCO conductors  and co-wound SQDs under:
	
	$$B_{\mathrm{op}} \le 3~\mathrm{T},\quad I_{\mathrm{op}} \le 600~\mathrm{A},\quad T_{\mathrm{op}}\ \text{variable}$$
	
	In practice, we use $T_{\mathrm{op}}\!\approx\!35$~K as a representative set-point (consistent with our load-line margin) and explore a current-density range by varying the copper cross-section. The operating current–density range studied here ($J_{\mathrm{op,cond}}=208\text{--}600~\mathrm{A\,mm^{-2}}$)
	matches the copper lamination options available for manufacturing. With a fixed operating current
	of $I_{\mathrm{op}}=600~\mathrm{A}$, stacking commercially available Cu tapes to different total
	stabilizer cross-sections $S_{\mathrm{Cu}}$ sets $J_{\mathrm{op,cond}} = I_{\mathrm{op}}/S_{\mathrm{Cu}}$
	within this range. Each conductor is paired with its dedicated co-wound SQD, and quenches are
	initiated by a centimeter-scale resistive heater for controlled triggering.
	\smallskip
	The 600 A available power supply constrains the width of the REBCO and copper tapes to 4 mm. The electrical insulation and thermal coupling between conductor and SQD are provided by a thin epoxy layer (about $100~\mu$m, constrained by handling and cool-down robustness). The heater geometry and placement are limited by the minimum quench energy and the width of the conductor.
	
	\subsection{Numerical Model on THEA}

	THEA (Thermal–Hydraulic–Electric Analysis) by Cryosoft® is a multiphysics simulation tool tailored to superconducting devices and cryogenic systems\cite{THEA24}. It solves transient, coupled problems along 1-D domains and has been recently used in studies adressing quench dynamics is fusion-relevant magnets:\cite{Chen2025QuenchCSLTS_TimeVarying,Cavallucci2023MultiphysicsQuench_ITERCS,Oh2024StabilityAnalysis_HTS_Fusion}. The features of the model are detailed below:
	
	\subsubsection{Conductor tape sub-model}
	For the stabilizer-only study, the THEA model contains a single \emph{thermal} component representing the conductor stack. Within this, we resolve three materials—REBCO, Hastelloy, and copper—each with temperature-dependent properties. The REBCO critical surface ($I_c(T,B,\theta)$ and $n$-value) is implemented through correlations inspired by the fit proposed in \cite{FleiterBallarino2014}, but re-optimized against our in-field and temperature-dependent measurements to match the tapes used here. The tape orientation is considered parallel to the external magnetic field i.e $B_{\mathrm{ext}} // (ab)$.
	
	\subsubsection{SQD sub-model}
	When the SQD is included, two additional thermal components are added: one for the SQD tape and one for the interposed glass-epoxy layer. Perfect thermal contact is assumed between the three thermal components, hence no thermal resistance is taken into account. Because the conductor and the SQD are electrically insulated, the electrical network is split into \emph{two independent electrical components} (one per tape) with infinite electrical resistance in between; each component carries its own transport current. The SQD critical properties are those of a REBCO tape, \emph{tuned} using literature-based relations that link a targeted $I_c$ degradation to the corresponding shift in $T_c$ \cite{Cayado2023}, consistent with the intentional deoxygenation approach adopted in this work. However, magnetic field's effect on deoxygenated tapes has not been modeled here, but future measurements on heat treated samples will are to be carried out to account for this dependence.

	\subsubsection{Quench trigger}
	Quenches are initiated by a localized heat pulse applied over a length equal to the minimum propagation zone, $L_{\mathrm{MPZ}}\simeq 35$\,mm, deduced by a trial-and-error process. The pulse power is set to $620~\mathrm{W\,m^{-1}}$ to match commercially available heaters and to reproduce realistic initiation conditions in the sample holder.
	
	\subsubsection{Mesh and domain}
	The computational domain is one-dimensional along the tape length. A uniform mesh of 2000 elements resolves a 2 m sample, which accommodates the heater zone, the voltage-gauge lengths, and sufficient run-out beyond the quench region to avoid boundary effects in the time window of interest.
	
	\subsubsection{Boundary conditions}
	As we consider an adiabatic quench in our case, the heat flux on both ends of the domain is fixed at zero. Concerning the electrical model, the total current is set at $I_{op,cond} + I_{op,SQD}$ but the "current-type" boundary conditions for each electrical component allow its distribution.
	\section{Results}
	Unless noted otherwise, results use a detection threshold $U_{\mathrm{threshold}}=0.1$\,V and a validation delay $t_{\mathrm{val}}\approx100$\,ms, consistent with our instrumentation's noise. Also, it is worth mentioning that the voltages here are calculated as the integral of the electric field along the whole length of each component respectively. The current dump phase is not modeled so that a longer-time thermal evolution can be observed (our geometry has very low inductance). \\
	In the results, $T_{det}$ is the temperature of the hotspot when the voltage crosses the threshold $U_{threshold}$ i.e at $t_{det}$, whereas $T_{dump}$ is the temperature after the validation of the quench, i.e at $t_{val}$.

	\subsection{Protection with Stabilizer Only}
	
		\begin{figure}[t]
		\centering

			\includegraphics[width=\linewidth]{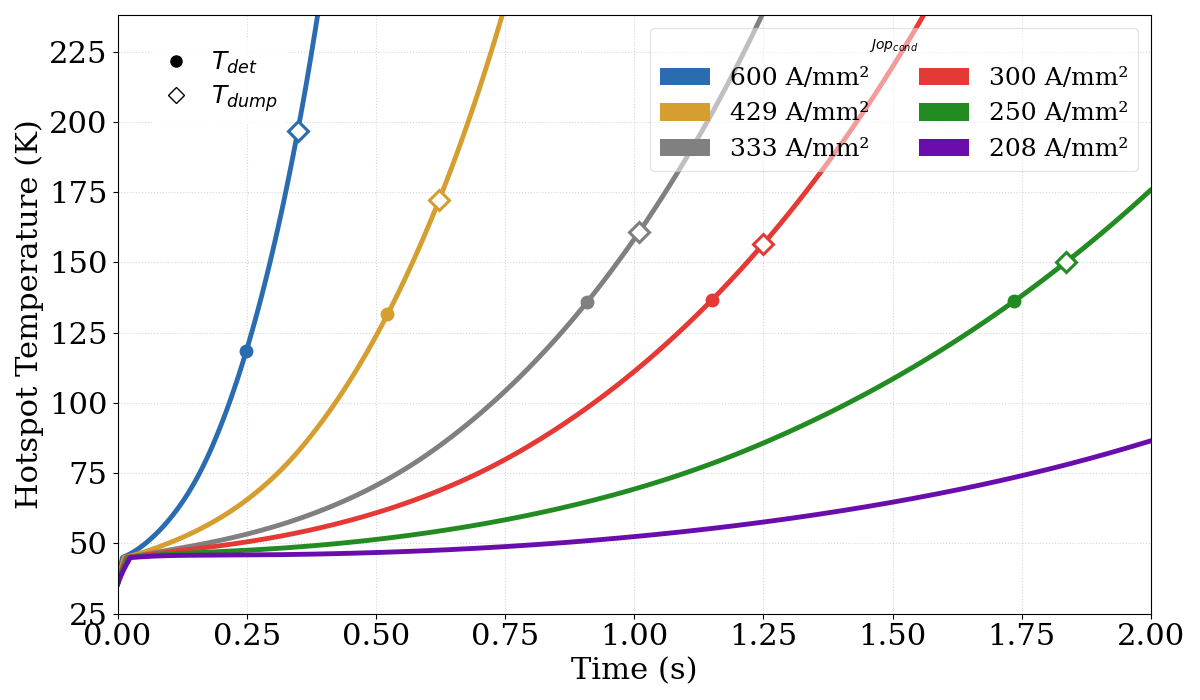}

			\includegraphics[width=\linewidth]{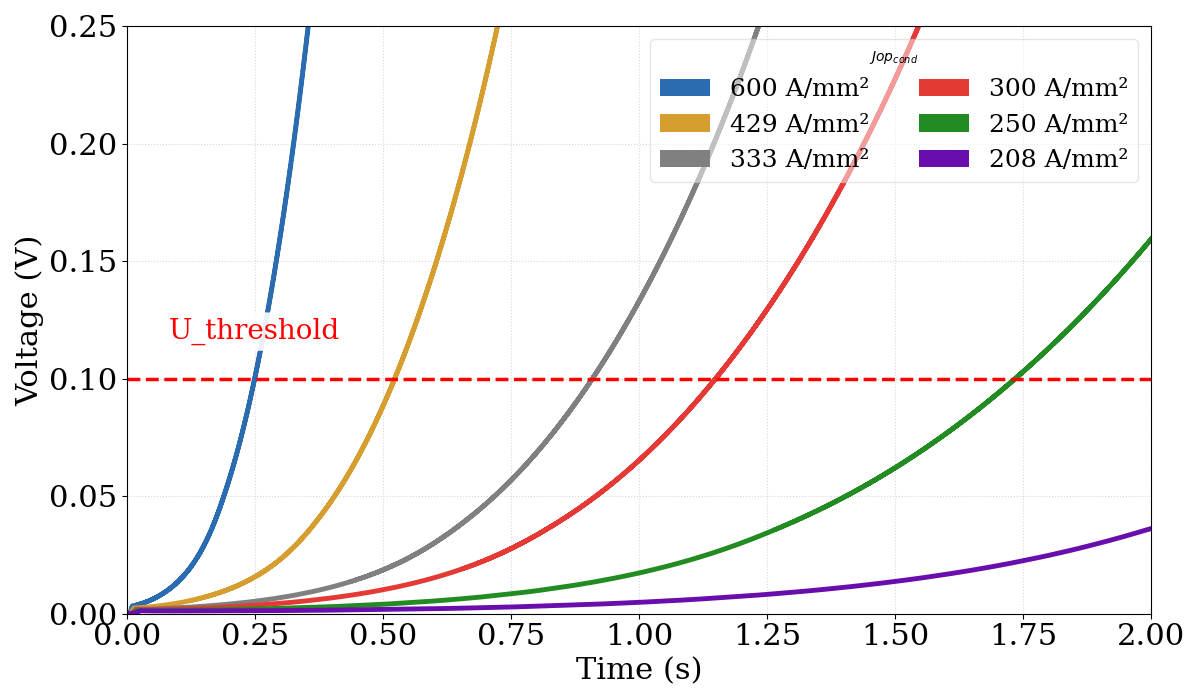}
		\caption{Stabilizer-only protection results. Evolution of the hotspot temperature (top) and conductor voltage (bottom) for different current densities}
		\label{fig:stabilizer_only_pair}
	\end{figure}
	
	Figure~\ref{fig:stabilizer_only_pair} shows the hotspot and voltage evolution
	for a series of conductors operating current densities
	$J_{\mathrm{op,cond}}$, obtained by varying the copper stabilizer area while
	keeping $I_{\mathrm{op}}=600$~A fixed. Here, it is important to separate the two phases of the quench : "detection" and "validation" and their corresponding temperatures. Indeed, As copper is reduced (larger $J_{\mathrm{op,cond}}$) the normal zone propagation velocity rises, which gives longer normal zones. Consequently, as the voltage is propotional to the normal zone length and inversely propotional to the copper cross section, the thinner the copper, the higher the measured voltage, the sooner the threshold is crossed. However, this lowers only the detection temperature $T_{det}$ because the quench is detected earlier, letting the Joule heating rise drastically during the validation time as the copper is too thin to spread it. On the other hand, as more copper is added (smaller $J_{\mathrm{op,cond}}$), the quench gets slower, allowing a late detection, but the heat capacity of the thick copper section is able to spread heat and minimize the Joule heating.\\
	However, we can conclude that the thicker copper section, although it slows down the quench, allows the hotspot to stay under a tolerable temperature of ~150 K. It is hence possible to design a "protectable" HTS stack-like conductor by optimizing stabilizer section, but if we want to reduce this temperature limit, an SQD can be used, which will be detailed in the following section.

	\subsection{Protection with SQD}
	Co-winding a lightly stabilized, electrically insulated, thermally coupled REBCO SQD creates a voltage channel that rises earlier and faster for the same disturbance, enabling earlier action at a cooler conductor hotspot. We keep the conductor's current fixed at 600 A, and we probe two key levers: the SQD operating current and the level of intentional degradation parameterized by $\alpha_{\mathrm{deg}}=(I_{c,0}-I_{c,\mathrm{deg}})/I_{c,0}$).

	\subsubsection{\textbf{Influence of the SQD operating current}}
	
\begin{figure}[!h]
		\centering

			\includegraphics[width=\linewidth]{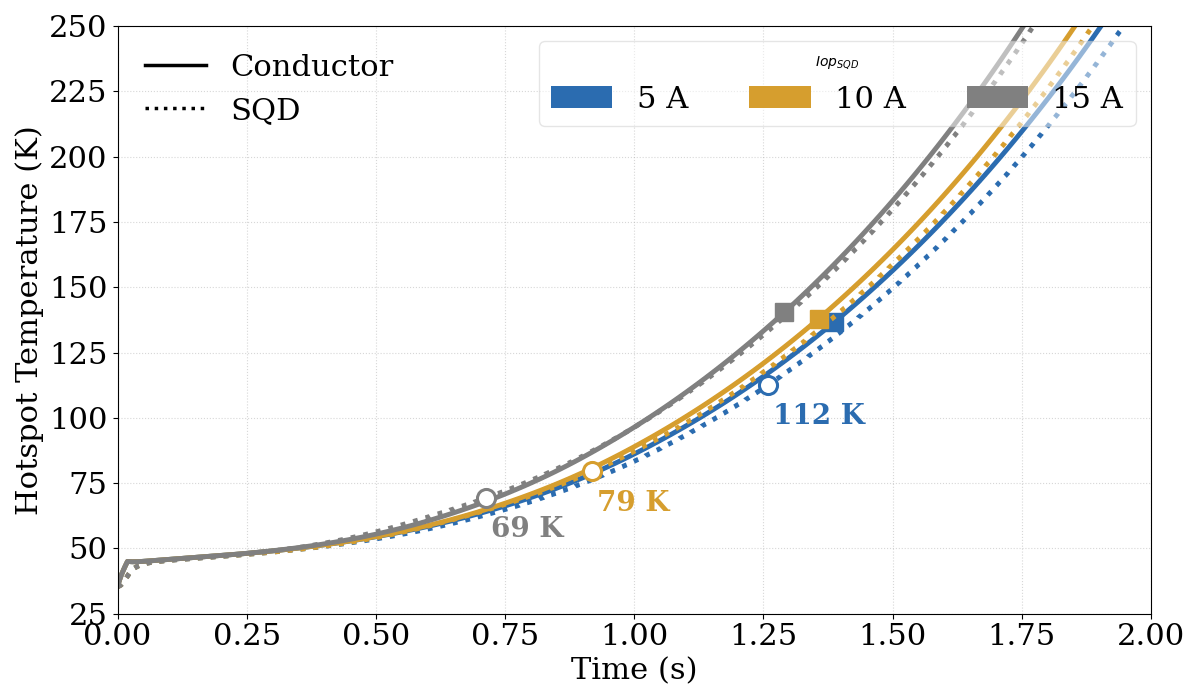}\vspace{0.35em}

				\includegraphics[width=\linewidth]{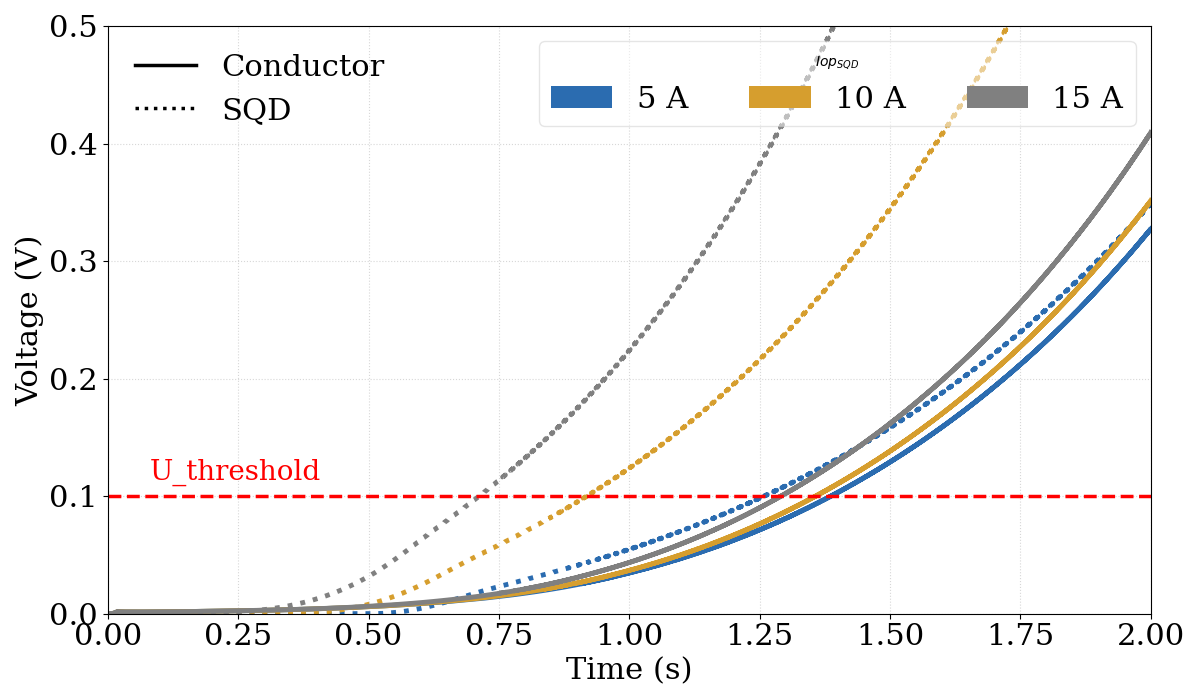}
				\caption{SQD-assisted protection: effect of SQD operating current. The square symbols are the temperatures of the conductor when its voltage $U_{det, cond}$ crosses the threshold. the round symbols are the temperatures of the conductor when the SQD voltage $U_{det, SQD} $crosses the threshold}
				\label{fig:sqd_iop_pair}
			\end{figure}

	Figure~\ref{fig:sqd_iop_pair} compares the baseline (no SQD) to SQD cases. Increasing $I_{\mathrm{op,SQD}}$ raises the SQD’s resistive response and shortens the time to threshold crossing, while the conductor’s thermal behavior remains essentially unchanged (negligible over-heating). We consider a degradation factor $\alpha_{\mathrm{deg}} = 80 \%$ and the corresponding critical current $Ic(T_{op},B_{op})=36 \ \mathrm{A}$. Hence the three currents $5 \ \mathrm{A}$ , $10 \ \mathrm{A}$ and $15 \ \mathrm{A}$ correspond to $\approx$ \ $14 \% $, $27 \%$, and $42 \%$ of $I_c$ respectively.
	
	From the plots, the threshold times and hotspot temperatures at detection are approximately:
	
		\[
	\begin{aligned}
		&\text{no SQD:}\ \ \ \ \ \ \ \ \ \ \ \ \ \ t_{\mathrm{det}}\!\approx\!1.35\text{ s},\ T_{\mathrm{det,cond}}\!\approx\!135\text{ K};\\
		&\text{SQD at 5 A:}\ \ \ \ \ \ \ \ \ \ t_{\mathrm{det}}\!\approx\!1.25\text{ s},\ T_{\mathrm{det,cond}}\!\approx\!112\text{ K};\\
		&\text{SQD at 10 A:}\ \ \ \ \ \ \ \ \ t_{\mathrm{det}}\!\approx\!0.95\text{ s},\ T_{\mathrm{det,cond}}\!\approx\!79\text{ K};\\
		&\text{SQD at 15 A:}\ \ \ \ \ \ \ \ \ t_{\mathrm{det}}\!\approx\!0.75\text{ s},\ T_{\mathrm{det,cond}}\!\approx\!69\text{ K}.
	\end{aligned}
	\]
	
	Thus, moving from 0 to 10–15\,A shortens detection by $\sim$0.4–0.6\,s and lowers $T_{\mathrm{HS}}$ at detection by $\sim$60–70\,K, creating a comfortable margin even after the validation delay. In practice, the choice of $I_{\mathrm{op,SQD}}/I_{\mathrm{c,SQD}}$ balances detection speed against SQD self-heating; the 27-42 \% range (10-15 A) provides a favorable compromise in our configuration.

\subsubsection{\textbf{Influence of the SQD’s degradation fraction $\alpha_{\mathrm{deg}}$}}%

\begin{figure}[!h]
	\centering
	%		\subfloat[Hotspot temperature vs.\ time.]{%
		\includegraphics[width=\linewidth]{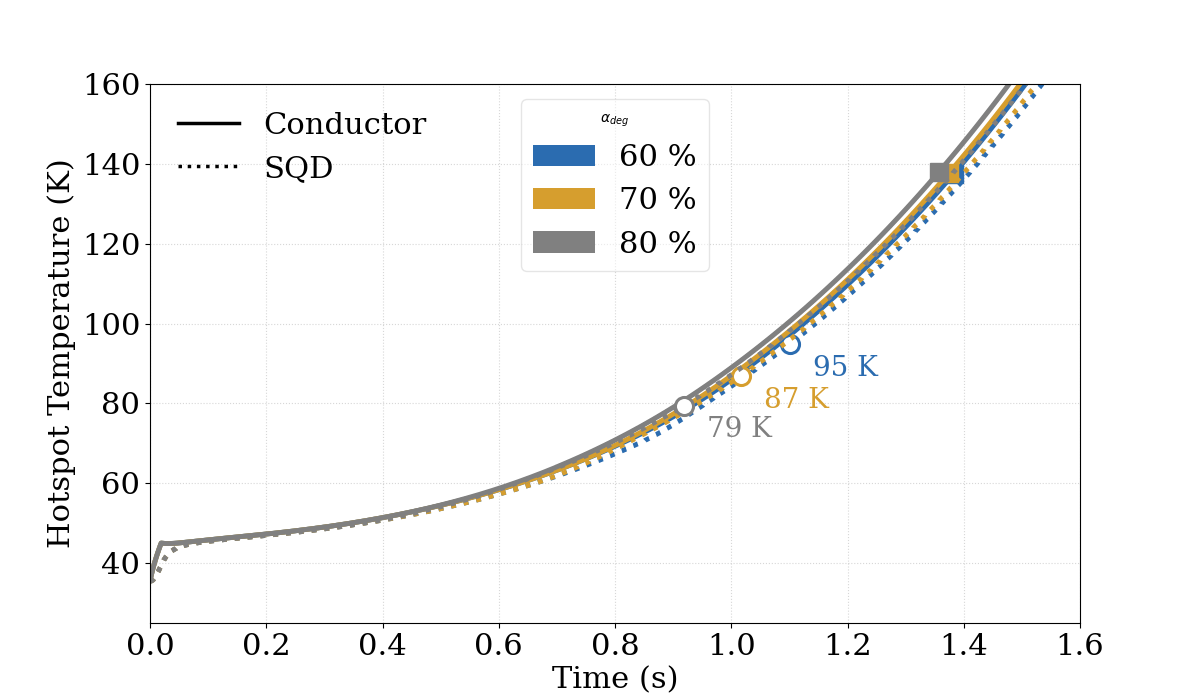}
		%		\subfloat[Voltages vs.\ time; $U_{\mathrm{threshold}}=0.1$\,V.]{%
			\includegraphics[width=\linewidth]{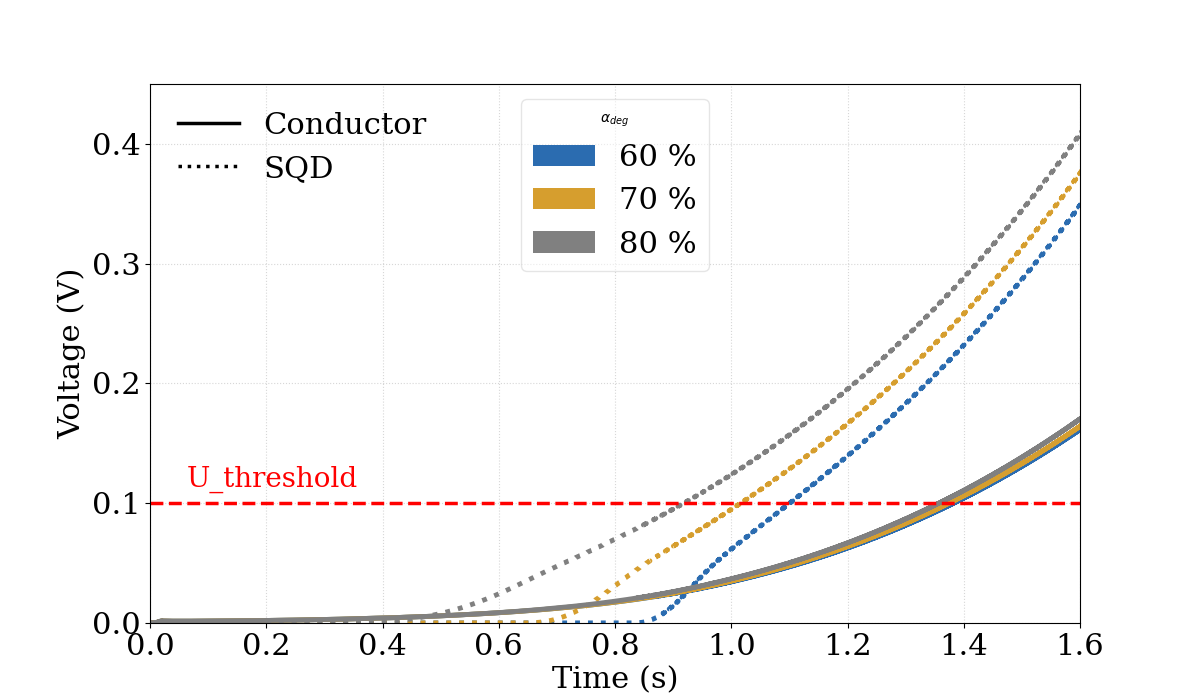}
			\caption{SQD-assisted protection: effect of intentional degradation. Larger $\alpha_{\mathrm{deg}}$ (stronger $I_c$/$T_c$ reduction) advances detection on the SQD channel and lowers $T_{\mathrm{HS}}$ at detection for the conductor.}
			\label{fig:sqd_deg_pair}
		\end{figure}

We next vary the SQD’s  degradation level for SQDs powered at 10 A (Figure \ref{fig:sqd_deg_pair}). Increasing $\alpha_{\mathrm{deg}}$ lowers the SQD’s $I_c$ (and associated $T_c$), shrinking the current-sharing window and making the SQD more sensitive to the conductor’s thermal rise. The outcome is an {earlier} voltage increase for the same disturbance and a cooler conductor at detection time $t_{det}$.

Analyzing the curves, the detection temperature decreases monotonically with degradation: about 95\,K (60\%), 87\,K (70\%), and 79\,K (80\%), with corresponding advances in $t_{\mathrm{det}}$. This confirms the design intention: tailoring the SQD’s critical surface (via controlled deoxygenation) is an efficient lever to improve detectability without relying on a substantially thin stabilizer in the main conductor.

	\section{Conclusion}
	
	We examined two protection strategies for insulated REBCO stacks. The copper stabilizer alone allows to protect the conductor in the case of a quench by keeping the hotspot temperatures below $150$ K in our case. Indeed, thick copper sections are able to compensate for the slow quench propagation by spreading and absorbing the heat due to the Joule effect in the normal zone. However, if one wishes to reduce the maximum tolerable temperatures during the quench, the use of an SQD is a promising choice. Thanks to its low electric stabilization, narrow current sharing phase and non-intrusive integration, it allows an earlier quench detection without risking overheating the conductor. A key element, however, is to find suitable tapes and degradation processes for the dedicated application, as the performances of these quench detectors depend on multiple factors like their operating current and degradation level.

	Finally, experimental validations of both strategies are being prepared in the H0 facility, covering multiple conductor geometries and SQD variants. These tests will help us estimate thermal contact resistances and confirm detection timing, hotspot temperatures, and integration choices.

\newpage
	% ---------- Acknowledgment ----------
	\section*{Acknowledgment}
	This work received support from the French State managed by the Agence Nationale de la Recherche under France~2030, grant ANR-24-EXSF-0006.
	
	% ---------- References ----------
	\bibliographystyle{IEEEtran}
	% Provide refs via refs.bib (e.g., Lu2021; Cayado2023; Pan2023; Peng2024; Bradford2024; THEAmanual; H0station; Poster; Thesis)
	\bibliography{refs}

\end{document}